**Design and Fabrication of a Microfluidic System with Nozzle/Diffuser Micropump and Viscosity**

This thesis submitted in partial fulfillment

of the requirements for the degree of

*Master of Science*

*in*

*Electronics and Communication Engineering by Research*

by

SUMANA BHATTACHARJEE

2018702011

sumana.bhattacharjee@research.iiit.ac.in

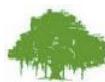

International Institute of Information Technology

Hyderabad - 500 032, INDIA

JULY 2021

International Institute of Information Technology

Hyderabad, India

**CERTIFICATE**

It is certified that the work contained in this thesis, titled "Microfluidic Sensors & Its Applications" by Sumana Bhattacharjee, has been carried out under my supervision and is not submitted elsewhere for a degree.

\_\_\_\_\_\_\_\_\_\_\_\_                                   \_\_\_\_\_\_\_\_\_\_\_\_\_\_\_\_\_\_\_\_\_\_\_\_\_\_\_

Date                                                          Advisor: Dr. Aftab M. Hussain

Assitant Professor

PatrIoT Lab, CVEST

IIIT Hyderabad, India

To Family and friends

# Acknowledgments

First and foremost I am very grateful to my advisor Dr. Aftab M. Hussain from the bottom of my heart, for his constant support and guidance, which made this research work possible. He has been there for every small doubt and patiently solved them. Not just by being a great guide as he is, he has been a friend too. For these three years, by motivating while I am stuck, by guiding in the right directions, by pushing me to try new things, by understanding when I need a break, by scolding when I needed, by giving life lessons, by listening to my endless chatters, he has always been there. I could not have asked for a better guide. Thank you very much, Sir.

Secondly, I am thankful to my all lab mates, for all the discussions, late-night chats, getting scared of Sir together, sharing food, which has helped me grow. I am grateful to my friends Mayank, Ruchi, & Deeksha who have been a constant support for all this time. I have to thank the entire IIIT community, each staff. They have created a beautiful environment so I could research without any disturbance. Their prompt responses to each issue, email has made this journey smooth.

Finally, the most important part, my family i.e. my mother and my sister, I thank you for being a part of my life. I am thankful to God for giving me such a blessing. They are the reason I could dare to dream and go after them to achieve them. Even when they did not understand a word I am talking about in my research, they were still there, listening patiently. Thank you.

# Abstract

In this thesis, we have discussed two microfluidic sensors, microfluidic micropump, and viscosity sensors, for biomedical and industrial applications. We have done simulations, mathematical analysis, and fabrication of the mentioned devices.

Micropumps are one of the most important parts of a microfluidic system. In particular, for biomedical applications such as Lab-on-Chip systems, micropumps are used to transport and manipulate test fluids in a controlled manner. In this work, a low-cost, structurally simple, piezoelectrically actuated micropump was simulated and fabricated using poly-dimethylsiloxane (PDMS). The channels in PDMS were fabricated using patterned SU-8 structures. The pump flow rate was measured to be 9.49 µL/min, 14.06 µL/min, 20.87 µL/min for applied voltages of 12 V, 14 V, 16 V respectively. Further, we report finite element analysis (FEA) simulation to confirm the operation of the micropump and compare favorably the experimentally obtained flowrate with the one predicted by simulation. By taking these flow rates as a reference, the chamber pressure was found to be 1.1 to 1.5 kPa from FEA simulations.

Viscosity measurement has wide-ranging applications from the oil industry to the pharmaceutical industry. However, measuring viscosity in real-time is not a facile process. This work provides an elaborate mathematical model and study of measurement of viscosity in real-time using pressure sensors. For a given flowrate, a change in liquid viscosity gives rise to a change in pressure difference across a particular section of the pipe. Hence, by recording the pressure change, viscosity can be calculated dynamically. Mathematical modeling as well as finite element analysis (FEA) modeling has been presented. A set of pressure sensors were placed at a fixed distance from each other to get the real-time pressure change. Knowing the flow rate in the channel, the viscosity has been calculated from the pressure difference. For the finite element analysis, the pressure sensors were placed 60 mm away from each other. The radius of the pipe was 19 mm. A different ratio of the mixture of water and glycerol was used to provide variable viscosity, which led to the variation in pressure-difference values.

# Contents

# List of the figures

# List of tables

*Chapter 1*

# Introduction

## 1.1   Motivation

In today's world of miniaturization, making affordable, small devices which are more efficient than bulky ones are desirable. To control less amount of fluids in a sophisticated manner, microfluidics plays a huge role. Pioneered by Standford university with the making of gas analyzing system having the concept of gas chromatography and heat sink [1][2], the research on microfluidics kept on growing. Many devices like flow sensors, micropumps, systems for chemical analyses, systems for separation of capillaries, etc. have been created since then [3]. But the main contribution of this field of research was in making biomedical devices like lab-on-chips. There are numerous advantages if the laboratory setup can be scaled down to μm range [4], like:

- A very little sample is needed. The reduction in volume happened from the factor of $10^3$ to $10^9$. The amount of fluid handled came down from 1mL to 1nL or 1pL.

- The system became very fast to analysis

- Efficient schemes for detection

- Systems became portable

- Affordable price and easy to mass-produce

Other than being very fast the lab-on-chip single-handedly can transfer samples, draw a precise amount of chemicals, mix reagents, heating, etc. within a few square centimeters [5]. It includes microchannels, micropumps, valves, sensors, etc. depending on the functionality of the device [6][7]. But making these small devices using silicon can be very expensive as we need to use silicon fabrication processing and need to use advanced facilities like cleanroom. So, adaptation to polymer molding techniques became more and more attractive. Nowadays using PDMS, PMMA, SU-8, etc. soft materials are widely used to make microfluidic systems [8][9][10]. These kinds of materials are easy to handle and flexible which makes them best suited for biomedical applications.

Over the past decades, there have been several pieces of research that have utilized the fundamental nature of these flexible materials [11]. PDMS is used much frequently due to its low-cost, transparent nature, hydrophobicity, stretchability, robustness, and biocompatibility[12]–[16]. Working with PDMS is pretty straightforward, after mixing with a curing agent, the liquid mixture is poured into the hard mold and heated at the proper temperature so the PDMS gets hardened but the flexibility remains [17][18]. PDMS was first reported to be used to get microfluidic devices in 1995 and there has been much research done in this area since then [19][20].

Further simplifying the process of manufacturing, nowadays, 3D printed devices are getting the attention of researchers. Without the need for cleanroom facilities, this kind of fabrication process brings down the fabrication cost in microfabrication even less [21]. 3D printing also allows using multiple materials, like hydrogels, polymers, metals, and ceramics [22]. There have been examples of making mixers [23], pumps, liquid separators [24], etc. using this technology. In the extension of my research work, I explored this area of research.

In this thesis, we have explored this ever-changing research area, microfluidics. We have discussed two devices here:

- Microfluidic micro-pump
- Microfluidic viscosity sensor

## 1.2   Contribution of this thesis

- The design of a working micropump, which has a very simple structure, an inlet working as a nozzle, an outlet working as a diffuser, and a circular chamber. Due to the trapezoidal structure, the liquid moves in uni-direction.

- The simulation and analysis of the micropump. The simulation shows how the trapezoidal structure helps in the working of the device.

- This novel structure of micropump gives motivation for future research work in this exciting area.

- In biomedical researches like disease diagnostics, health monitoring, etc. can get a significant amount of boost from this research.

- A novel way of sensing the real-time viscosity of the liquid flow was discussed.

- Using just a pressure sensor, one of the very complex properties of liquid, viscosity can be measured. That process has been discussed.

- This sensor can give a tremendous outcome in the oil, health, etc. industries.

## 1.3  Thesis organization

The rest of the thesis organized as follow:

- Chapter 2 talks in detail about the micropump. The related work, design of the pump, simulation results, experimental setup, and results have been discussed in this chapter.

- In chapter 3, the details of the viscosity sensor have been discussed. The related work, design of the sensor, simulation result, mathematical modeling, the comparison between the simulation and mathematics has been discussed here.

- In this last chapter, i.e. chapter 4, we concluded the entire work. Also, what can be done in the future has been discussed in the concluding chapter.

*Chapter 2*

# Simulation and Fabrication of Piezoelectrically Actuated Nozzle/Diffuser Micropump

## 2.1 Background

With the miniaturization of electronic devices and the advent of next-generation wearable electronics [25]–[28][29]–[33], the concept of a miniature laboratory for instantly analyzing biofluids has become possible. Micropumps play one of the most important roles in the making of microelectromechanical systems (MEMS) based microfluidic systems [34]. Micropumps have significant applications in lab-on-a-chip systems as well as embedded medical devices to exert bodily fluids, insert medicine in the body and help liquid flow. While there are many designs for microfluidic pumps in the literature, one of the simplest designs is the valveless nozzle-diffuser pump which employs a central chamber connected to the inlet and outlet through two flow diodes (Figure 2.1a). When the central chamber is repeatedly pressurized, the fluid flows preferentially from inlet to outlet. Generally, the chamber is pressurized using a flexible piezoelectric disc. The diffuser/nozzle structure of inlet and outlet channels allows more flow in one direction than the other. When the chamber expands, inlet behaves as diffuser and outlet behave as the nozzle. As a result, more flow is obtained through the inlet into the chamber than through outlet out of the chamber. When the chamber contracts, inlet behaves as nozzle and outlet behave as the diffuser. Thus, more flow is obtained through the outlet out of the chamber. The principle of operation of the micropump is shown in Figure 2.1b. We have used flexible PDMS as the pump material, which is biocompatible, and transparent, which is best suited for biomedical applications. We have done mathematical analysis and created a bridge between pressure in the chamber and voltage applied to the piezoelectric device, which is never done in literature before.

## 2.2   Related work

The first time micropump was developed in 1980 by Smits at Stanford University, this pump had a microvalves-based design [35]. After that, there have been many designs using membrane have been shown in the literature [36]–[40][41]. But due to leakage, the microvalves got replaced by a nozzle/diffuser [42]. The principle concept of nozzle/diffuser structure is by transforming kinetic energy i.e. flow velocity to potential energy i.e. pressure. The recovery of pressure has a high impact in the direction of the diffuser but not in the nozzle direction, this gives a pressure difference between the diffuser and nozzle, as a result, the fluid starts to move in the direction of the diffuser instead of the nozzle [3].

Due to the simplicity of this concept, there have been numerous examples in literature and the improvement kept continuing. In one of the early designs of this kind of pump shown by Olsson et al., there were two chambers used in the system, both of them with a vibrating diaphragm to excite the liquid in phase and out of phase [43]. In another literature, Jiang et al. investigated the change in flow rate due to the change in conical angle of the micro valveless pump, the had experimented with the angles of 5°, 7.5°, and 10° [44]. In another paper, Yun et al. used surfaced tension-based electrowetting to create a low-power micropump, where surface tension of mercury droplet and electrolyte as actuation energy was used to form the micropump [45]. A thermal-bubble-actuated micropump with nozzle/diffuser floe regulator was demonstrated by Tsai and Lin, here the actuation was done using the nucleating and collapsing of the thermal bubble using a resistive heater and a nozzle/diffuser flow controller [46].

A simple thermopneumatic PDMS and ITO-based micropump were demonstrated by this same group later [47]. Another multi-stacked PDMS-based thermopneumatic micropump was fabricated on a glass substrate by Mamanee et al. This peristaltic motion-controlled pump had a simple structure with one inlet, one outlet, three actuation chambers, and three heaters [48]. Ha and co-authors presented glass and PDMS-based fully integrated thermoneumatic nozzle/diffuser micropump having microscale check valves. This three-layer PDMS-based pump has Cr/Au heater fabricated on the chip. The design is made in such a way that the heater chip can be used over and over but the PDMS section can be disposable [49]. Another thermoneumatic micropump with three layers of PDMS stacking was demonstrated by Yang and Lin. In this structure, due to the flow rectification effect, the valveless nozzle/diffuser was driven. By Joule heating of the embedded heaters, the actuation stroke transfers the working liquid to the vapor inside the evaporation chamber [50].

A piezoelectrically actuated, structurally simple, low-cost disposable diffuser micropump was fabricated by Kim et al. using PDMS as base material [51]. Another piezoelectrical actuation-based micropump was demonstrated by Rao et al. for drug delivery, having four main elements, i.e., flexible PDMS-based membrane, a piezoelectric layer, electrodes, and reservoir [52]. Another valveless, planar, nozzle/diffuser microfluidic pump was demonstrated by Xia et al., whose actuator material was made of poly(vinylidene fluoride-trifluoroethylene) [P(VDF-TrFE)], this material has a high electrostrictive strain, approximately 5-7% and its highly elastic, this properties helped in actuator configuration [53]. Yang et al. in this paper discussed the numerical performances of nozzle/diffuser micropump in both series and parallel combinations [54].

An electromagnetic actuator-driven single chamber, valveless, nozzle/diffuser micropump was fabricated by Zhou and Amirouche. This PDMS-based pump had a nozzle/diffuser structure to control the fluid flow and embedded a thin magnetic membrane to pressure regulation [55]. In another paper, Said et al. demonstrated a hybrid membrane-based, electromagnetically actuated valveless micropump. This structure is made by a combination of magnetic polymer (composite of PDMS and NdFeB), and a bulk permanent magnet to generate a strong magnetic field as well as maintains flexibility [56].

## 2.3   Structure and Principle

The structure of the micropump is shown in figure 2.1. The pump chamber and channels are made out of molded PDMS. The PDMS is bonded with a glass substrate using the surface activation method. A piezoelectric buzzer is attached to the top of the PDMS chamber. The piezoelectric buzzer acts as the actuator for the pump. Inlet and outlet holes are mechanically drilled in the PDMS.  The micropump is a valveless pump. The diffuser/nozzle structure of inlet and outlet channels allows more flow in one direction than the other. When the chamber expands, the inlet chamber behaves as a diffuser and the outlet behaves as a nozzle. As a result, more flow is obtained through the inlet into the chamber than through the outlet out of the chamber. When the chamber contracts, the inlet behaves as a nozzle and the outlet behaves as a diffuser. Thus, more flow is obtained through the outlet out of the chamber. The principle of operation of the micropump is shown in figure 2.1.

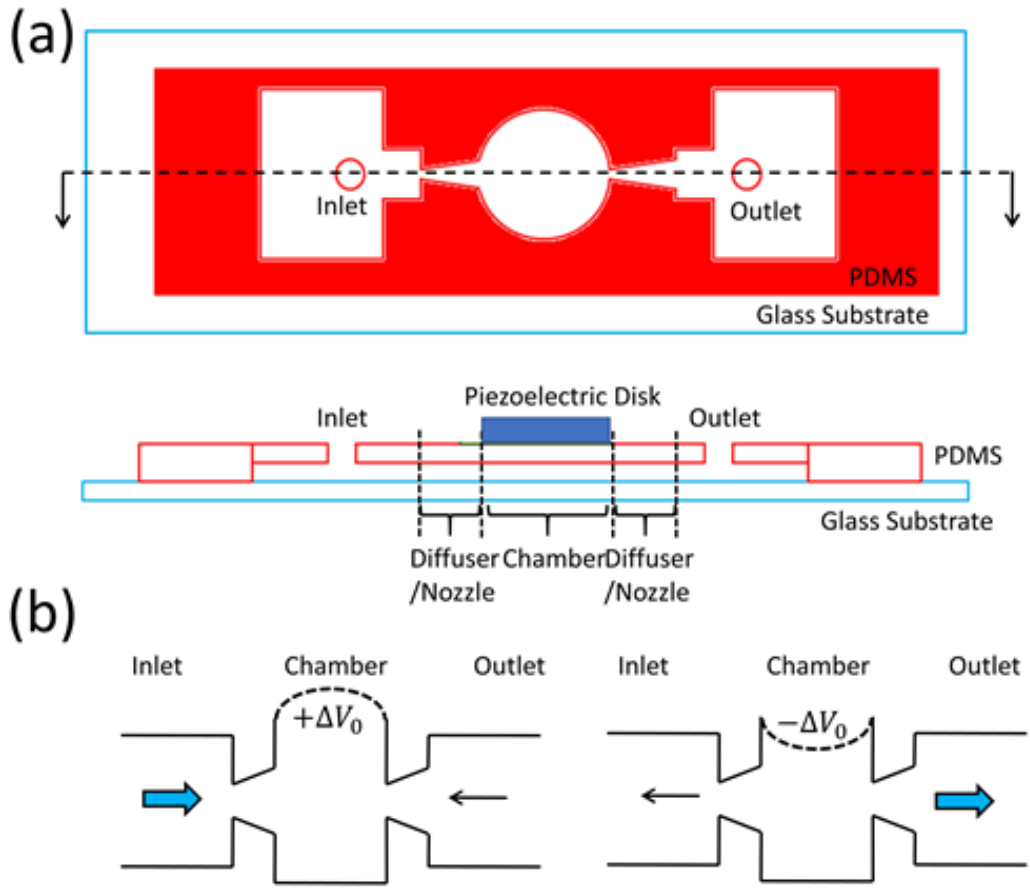

*Figure 2.1. (a) Structure of the diffuser/nozzle micropump. (b) Principle of operation of the micropump.*

## 2.4   Simulation Analysis

The micropump was simulated using COMSOLTM Multiphysics by applying different pressures on the chamber assuming that the pressure is uniformly distributed on the circular shape of the chamber. The pressure distribution and fluid flow from a trapezoidal diffuser/nozzle the structure was simulated for various applied pressure differences. As expected from theory, for the same applied pressure difference, the fluid flow was found to be less in the case of nozzle and more in the case of diffuser configuration. This difference in fluid flow is the cause of the net fluid flow from the inlet to the outlet in the micropump design (Figure 2.1b). The pressure distribution for the trapezoidal structure as diffuser and nozzle is shown in Figure 2.2. The length of the trapezoid structure was 1500 μm, while the width of the two ends was 40 μm and 200 μm. For an applied pressure difference of 1000 Pa, the flow rate was found to be 53.7 µL/min in the case of the diffuser and 45.7 µL/min in case of the nozzle configuration. The difference between these is proportional to the net flow rate in a given actuation cycle.

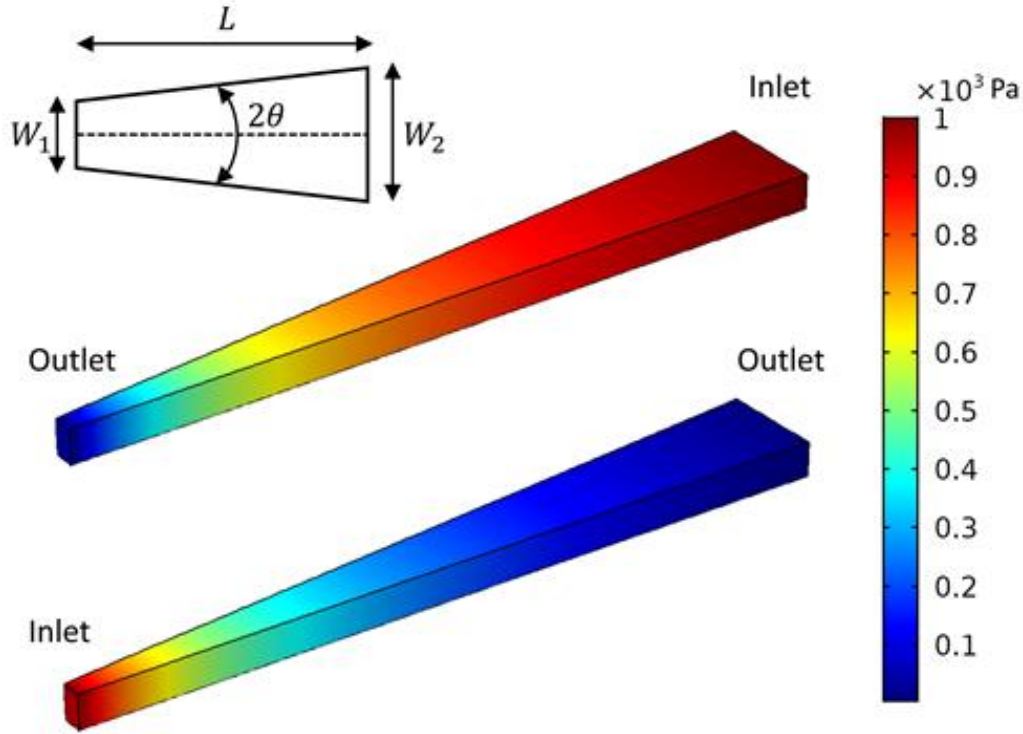

*Figure 2.2  Results of FEA simulations for pressure distribution in a trapezoidal microfluidic channel with L=1500 μm, W2=200 μm, W1=40 μm, as diffuser and nozzle.*

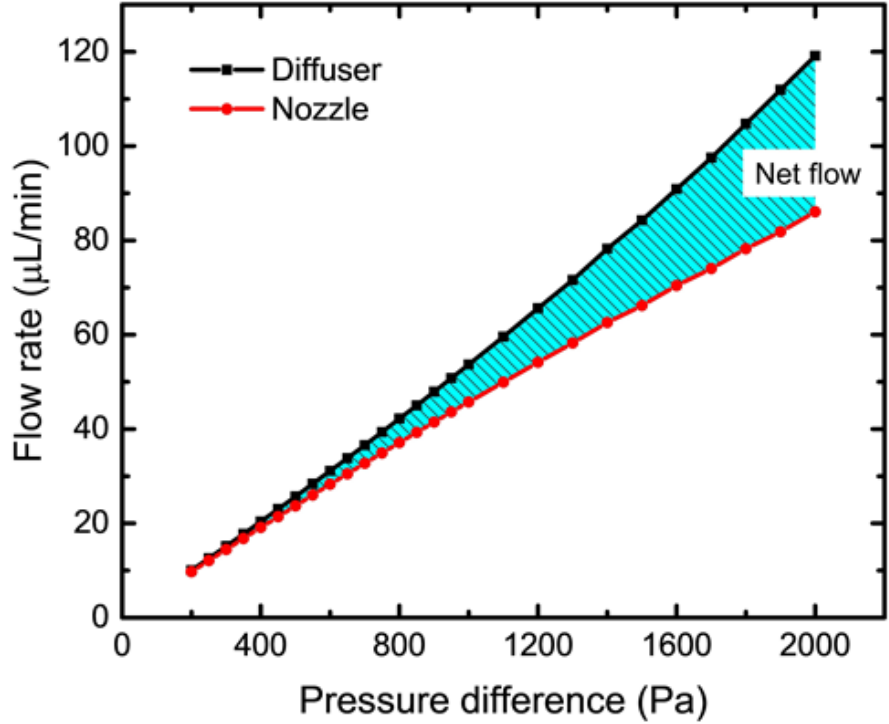

*Figure 2.3. Results of FEA simulations for flow rate as a in a diffuser and nozzle structure as a function of applied pressure. Simulated for L=1500 µm, W2=200 µm, W1=40 µm*

We simulated the net flow rate for various applied pressure differences by observing the difference in the flow rates for the diffuser and nozzle structures. The change in flow rate for a nozzle/diffuser pump concerning pressure is shown in Figure 2.3. The difference between the two flows (highlighted in Cyan) gives the net flow in the case of the nozzle/diffuser micropump.

## 2.5    Fabrication And Process

### 2.5.1    Fabrication of SU-8 mold

A thick layer of SU-8 was patterned on a Silicon wafer to make the PDMS layer consisting of the chamber, channels, and inlet-outlet reservoirs. The Silicon wafer was first cleaned with Acetone, IPA, and DI water. The wafer was subjected to dehydration bake at 110 °C for 5 minutes. SU-8 2025 was spun on the wafer using the recipe given in table 2.1.

| Speed (rpm) | Acceleration (rpm/sec) | Time (sec) |
|:-----------:|:----------------------:|:----------:|
| 500 | 100 | 10 |
| 1250 | 300 | 40 |
| 0 | -1000 | 0 |

*Table 2.1: Recipe used for spinning thick SU-8 layer*

The spun wafer was pre-baked at 65 °C for 1 minute and 95 °C for 6 minutes. The wafer was allowed to cool and then exposed using i-line UV. The exposure dose was set at 250 mJ/cm2. The wafer was then post-baked at 65 °C for 1 minute and 95 °C for 6 minutes. The wafer was immersed in a Microchem SU-8 Developer bath for 6 minutes for development.

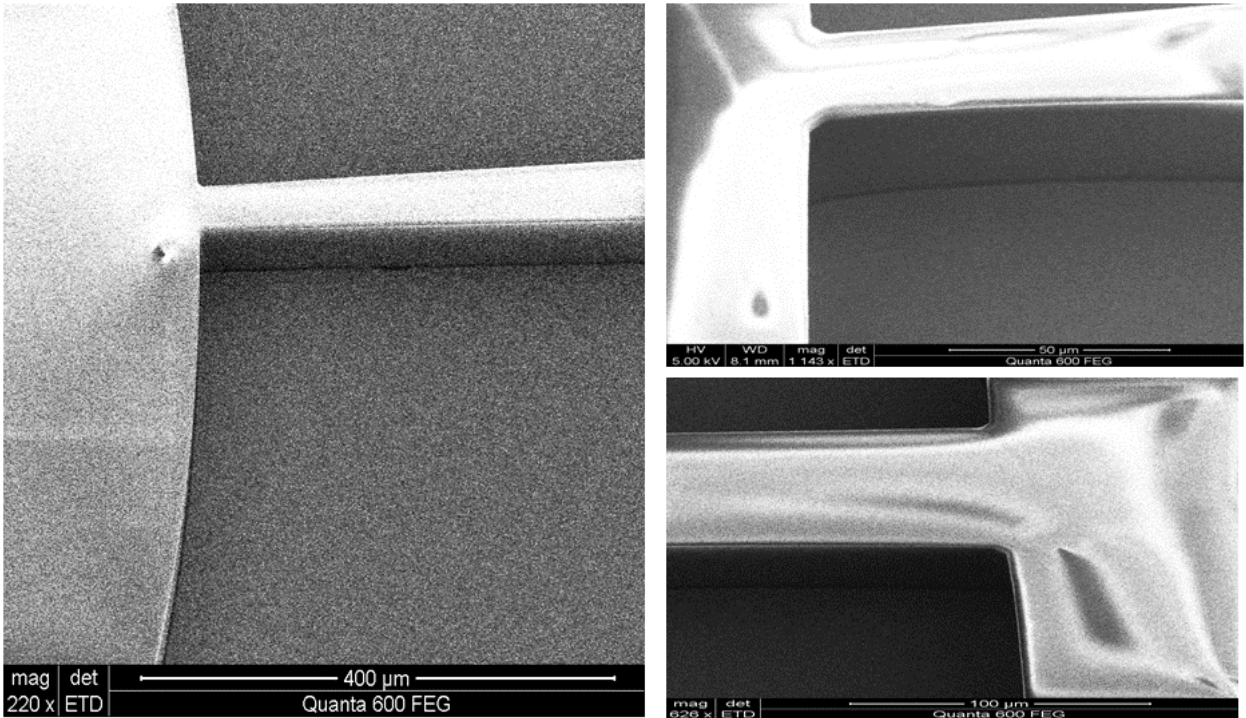

*Figure 2.4. SEM images of pattered SU-8 layer*

After development, the thickness of the SU-8 layer was measured using a profilometer. The thickness was found to be 70 µm. Figure 2.4 shows the SEM images of the patterned SU-8 layer. The SEM images show that SU-8 side-walls obtained using the above recipe were smooth and straight.

### 2.5.1.1 Mask Design

The mask used for patterning the SU-8 layer was a 5" dark field mask. Since SU-8 is a negative tone photoresist, to obtain free-standing structures of SU-8 on Silicon, a dark field mask was used. The micropump design on the mask is as shown in figure 2.5. The mask design consists of a chamber, diffuser/nozzle structure for inlet and outlet, inlet and outlet reservoirs. The mask contains micropumps of different diffuser/nozzle widths and chamber sizes:

- Chamber: 2 mm and 3 mm diameter.

- W1: 20, 40, 60 μm.

- W2: 60, 100, 200, 300 μm.

The complete mask design is shown in figure 2.5.

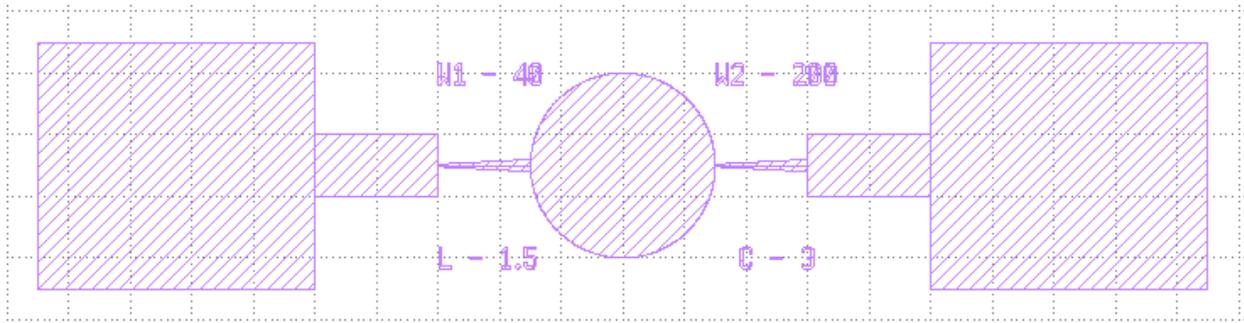

*Figure 2.5. Micropump Design*

## 2.5.2 Fabrication of PDMS layer

The PDMS prepolymer mixture was created by mixing Sylgard 184 and curing agent in the ratio of 10:1 by weight. The mixture was then poured into Aluminum weigh boats containing the SU-8 molds of the micropump as shown in figure 2.6. Typically, 4 grams of Sylgard 184 and 0.45 grams of curing agent were mixed to be poured into the weigh boat. The mixture was then degassed by exposing it to a vacuum in a vacuum oven. The mixture was simultaneously cured in the oven at 110 °C for 30 minutes. The PDMS layer thus formed was peeled off and cut into pieces with a single micropump design.

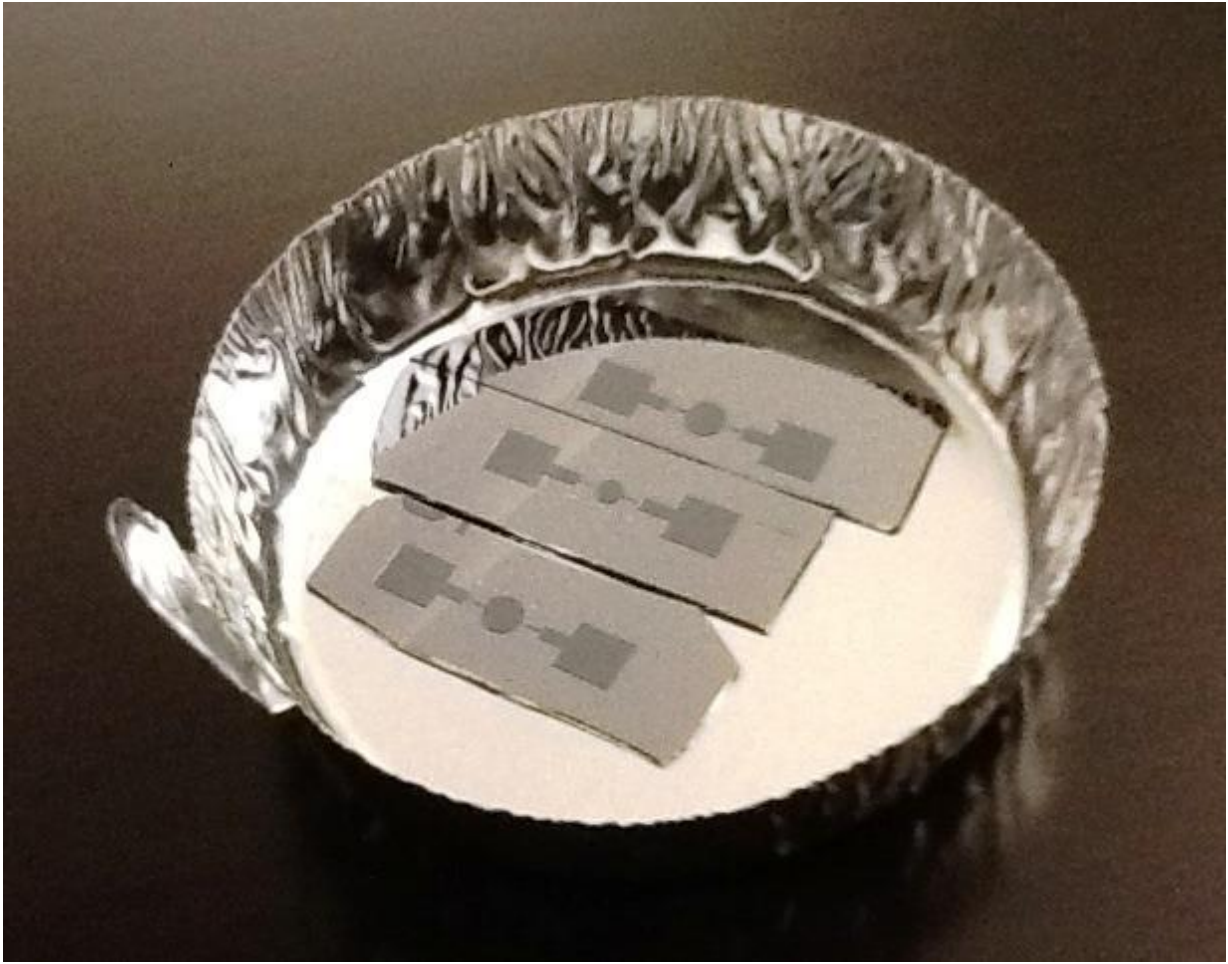

*Figure 2.6. Aluminum weigh boat containing SU-8 molds of micropumps*

### 2.5.3  Bonding PDMS layer to the glass substrate

The PDMS layer was bonded to the glass substrate by the surface activation method. Both the PDMS surface and the glass surface were exposed to Oxygen plasma in an RIE tool. Care had to be taken that both the glass surface and PDMS surface are completely clean. Before exposure to Oxygen plasma, the PDMS surface and glass substrate was cleaned using Acetone, IPA, and DI water. The recipe used for Oxygen plasma is given in table 2.2.

| Oxygen | 25 sccm |
|---|---|
| RF Power | 10 W |
| ICP Power | 100 W |
| Pressure | 70 mTorr |
| Time | 40 Sec |

*Table 2.2: Oxygen plasma recipe for surface activation of PDMS and glass substrate*

The micropump was quickly bonded after exposure. The activated PDMS surface was firmly pressed against the glass surface to affect the bonding. Figure 2.7 shows an assembled micropump.

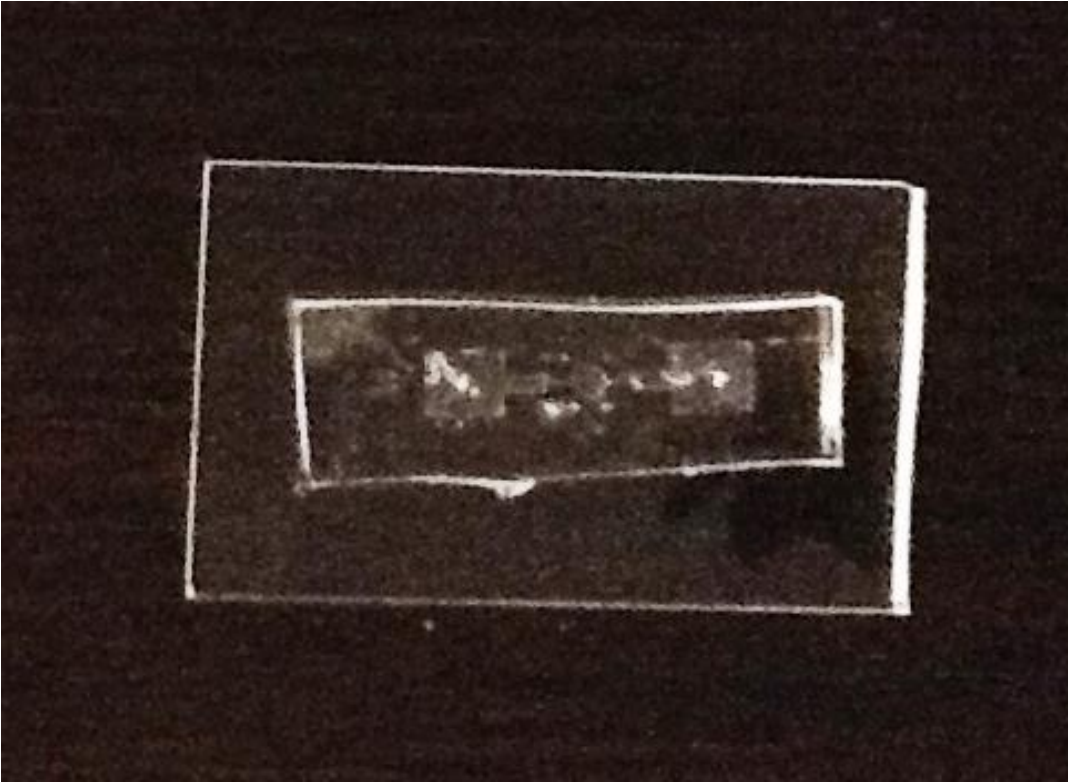

*Figure 2.7.An assembled micropump*

### 2.5.4 Piezoelectric buzzer

The pump was designed to be actuated using a piezoelectric buzzer. A Radio Shack Mini 12 VDC Electric Buzzer was used for the purpose. Contact rings were attached to the buzzer to fit on top of the PDMS chamber. The contact rings were cut out of a 2 mm thick solid PMMA sheet. The rings were glued together using Chloroform. The rings were attached to the buzzer using double-sided carbon tape. Figure 2.8 shows the completed piezoelectric buzzer assembly.

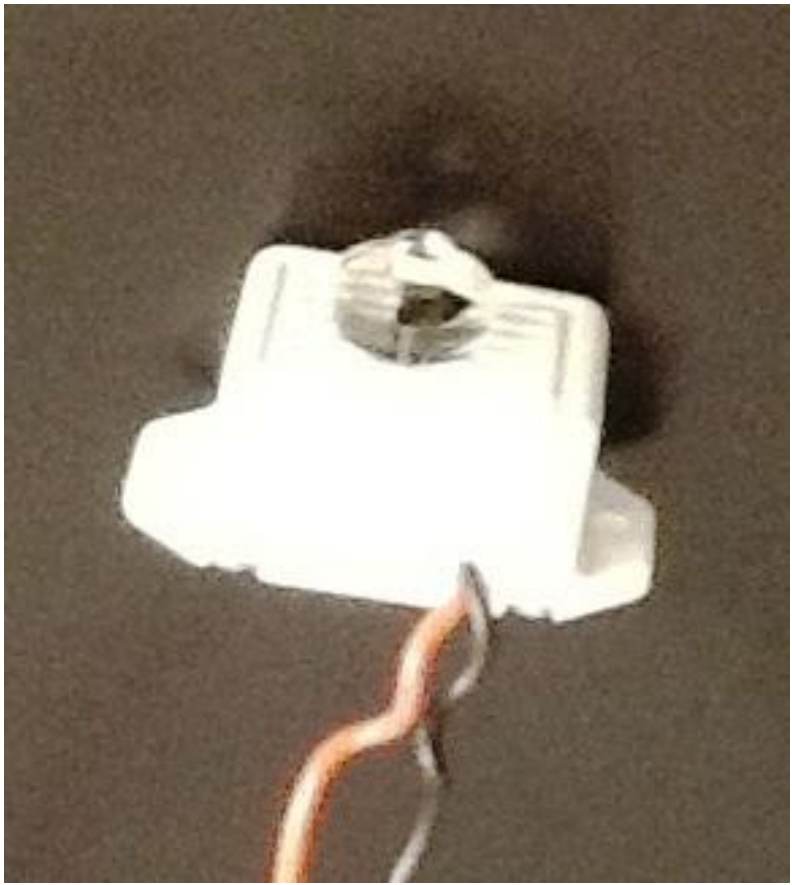

*Figure 2.8.Completed piezoelectric buzzer assembly*

### 2.5.5 Final Assembly

The final assembly was made from solid PMMA. A base platform was made to support the glass

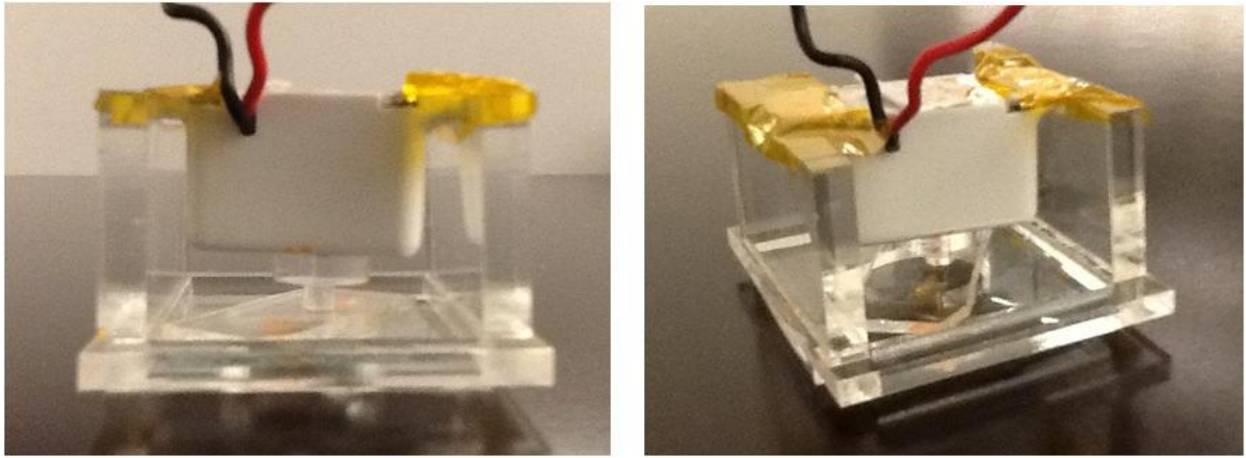

*Figure 2.91 Final micropump assembly*

substrate with the PDMS pump. A support structure was made for the piezoelectric buzzer, such

that the contact ring exactly touches the top of the PDMS chamber. The PMMA pieces were glued

together using Chloroform. Figure 2.9 shows the final assembly of the micropump. Figure 2.10

shows an assembled micropump with a colored die in the chamber and inlet/outlet reservoirs. Two small cavities were made in the PDMS layer using syringes to serve as inlets and outlets of the fluid.

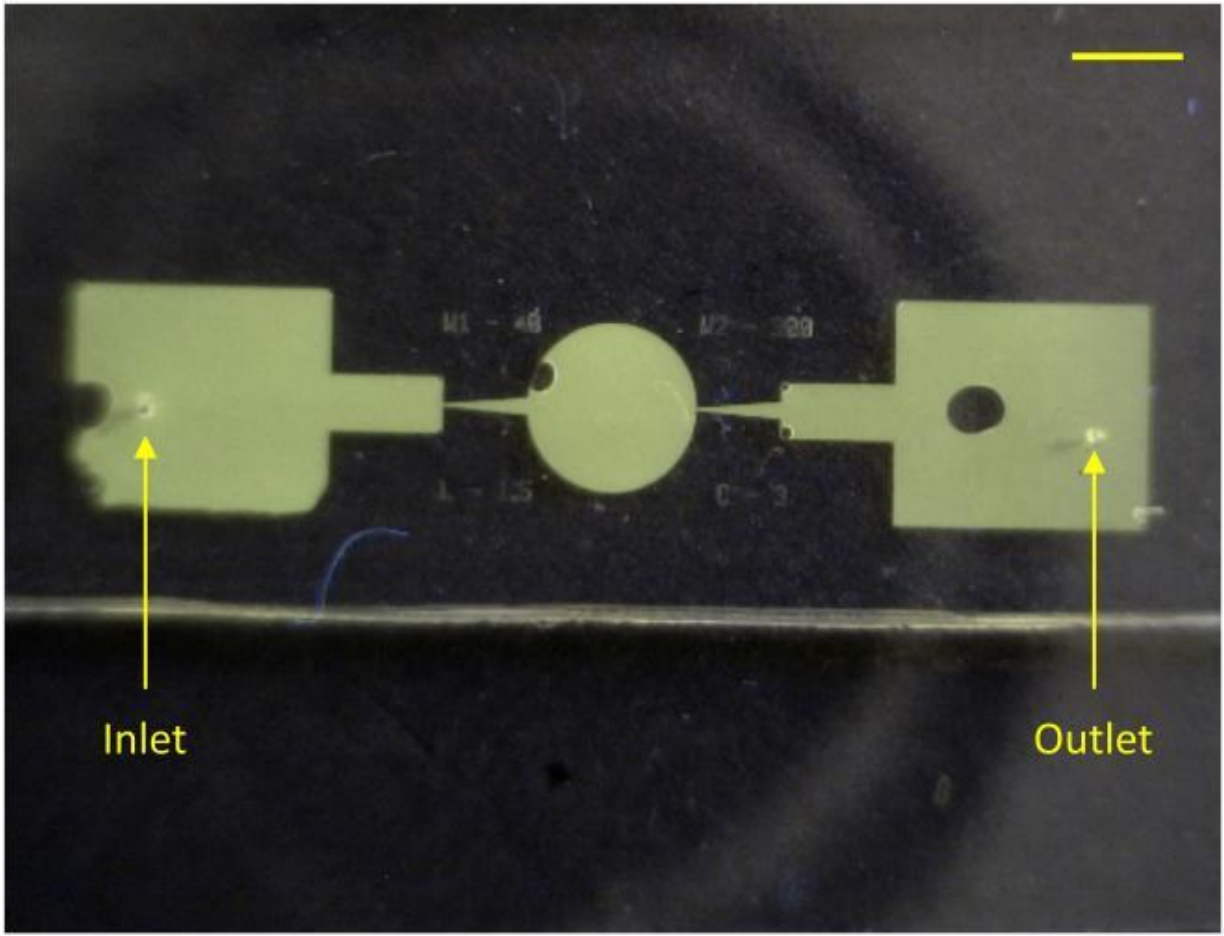

*Figure 2.10. Optical image of the final fabricated micropump with inlet and outlet syringe cavities, filled completely with colored die. Scale bar is 2 mm.*

## 2.6   Results Obtained

The micropump was tested at various stages of assembly for its functionality. Figure 2.11 shows the initial micropump assembly being tested manually; using a hand-held piezoelectric buzzer, and using the final support structure. The micropump showed expected behavior at every stage.

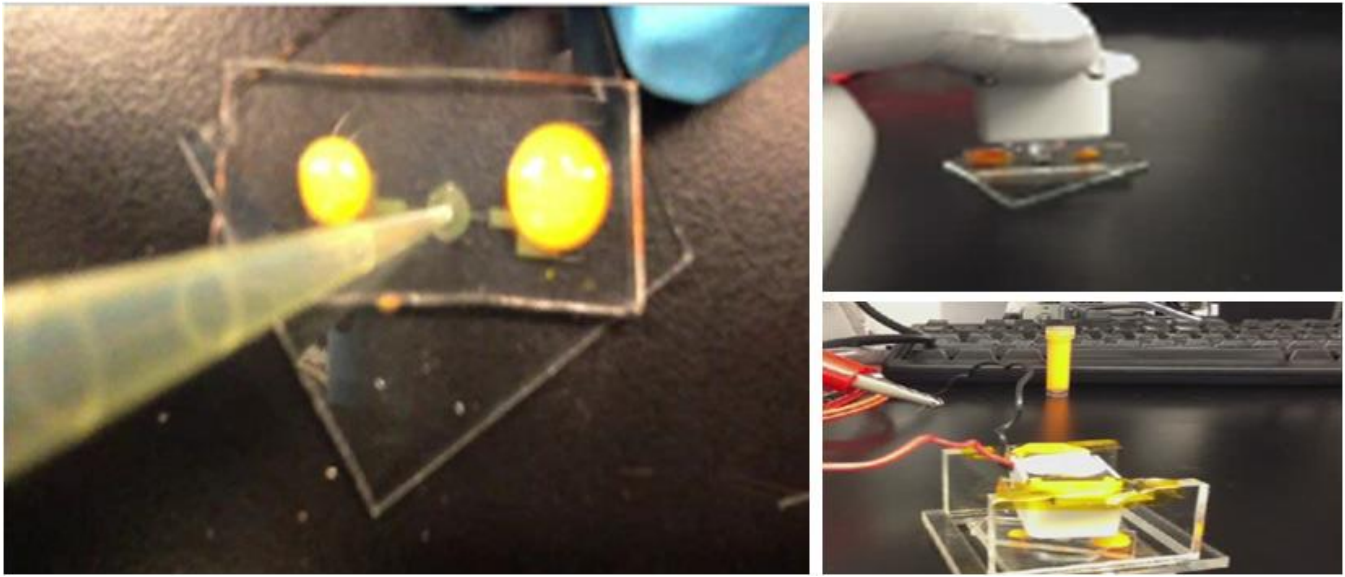

*Figure 2.11 Testing the micropump at various stages of fabrication*

The final assembly of the micropump was used to measure the flow rate for various applied voltages. The flow rate was determined by measuring the time taken to transport a known volume of liquid from inlet to outlet. Table 2.3 shows the results obtained from the measurement. These measurements were done for a micropump with a chamber size of 3 mm. The diffuser inlet width, its length, and its outlet width were 40 μm, 1500 μm, and 200 μm respectively.

| Voltage (V) | Current (mA) | Volume (mL) | Time | Flow rate (μL/min) |
|:---:|:---:|:---:|:---:|:---:|
| 12 | 12 | 0.04 | 4 min, 13 sec | 9.49 |
| 14 | 14 | 0.03 | 2 min, 08 sec | 14.06 |
| 16 | 16 | 0.04 | 1 min, 55 sec | 20.87 |

*Table 2.3: Flow rate measurement results*

In general, it was found that the pump performance was highly dependent on the dimensions of the nozzle/diffuser (L, W1, and W2). This was also observed from simulations. With the variation of W1 and W2, the net flow rate varies for a given applied pressure difference. Given the dimensions of the nozzle/diffuser assembly and assuming the same pressure difference during positive and negative actuation, the efficiency of the nozzle/diffuser structure, η, can be calculated from the following equation [11, 12]:

$$\eta = \left(\frac{19}{20}\right)\left(\frac{W_2}{W_1}\right)^{0.34}$$

(2.1)

From the above equation, the value of η was found to be 1.642, for $W_1$=40 μm and $W_2$=200 μm. Now, the net volume flow rate of the nozzle/diffuser structure has been derived as [13]:

$$Q = 2\,\Delta V\,f\left(\frac{\eta^{\frac{1}{2}} - 1}{\eta^{\frac{1}{2}} + 1}\right) = 2\,\Delta V\,f\,C$$

(2.2)

Where Q is the net flow rate, ΔV is the volume change per each actuation cycle, $f$ is the frequency of actuation and C is the rectification factor calculated as,

$$C = \left( \frac{\eta^{\frac{1}{2}} - 1}{\eta^{\frac{1}{2}} + 1} \right)$$

(2.3)

Thus, C is calculated to be 12.33% for the fabricated micropump with $W_1$=40 µm and $W_2$=200 µm. The term $\Delta V$ in the above equation depends on the deflection of the diaphragm, which in turn depends on the applied voltage in the case of a piezoelectric actuator. Hence, because C is a constant for given micropump dimensions, the net flow rate Q only depends on the applied pressure difference. This is also evident from the FEA analysis reported earlier.

To verify the simulation and experimental analysis, we have plotted the net flow rate and pressure difference for the simulation results (as we can see in Fig 2.12 blue line). In the case of the piezoelectric actuator, the independent variable is the applied voltage which was mapped to applied pressure using a constant multiplier because these quantities are linearly dependent. The experimental results were obtained by comparing applied voltage with flow rate. Figure 2.12 shows experimental and simulation results for the fabricated micropump with L=1500 µm, $W_2$=200 µm, $W_1$=40 µm. By fitting the experimental results to the simulation, the scaling factor for the applied voltage was found to be 97.24 Pa/V. Thus, the pressure in the micropump chamber was found to be in the range of 1.1 to 1.5 kPa (Figure 2.12).

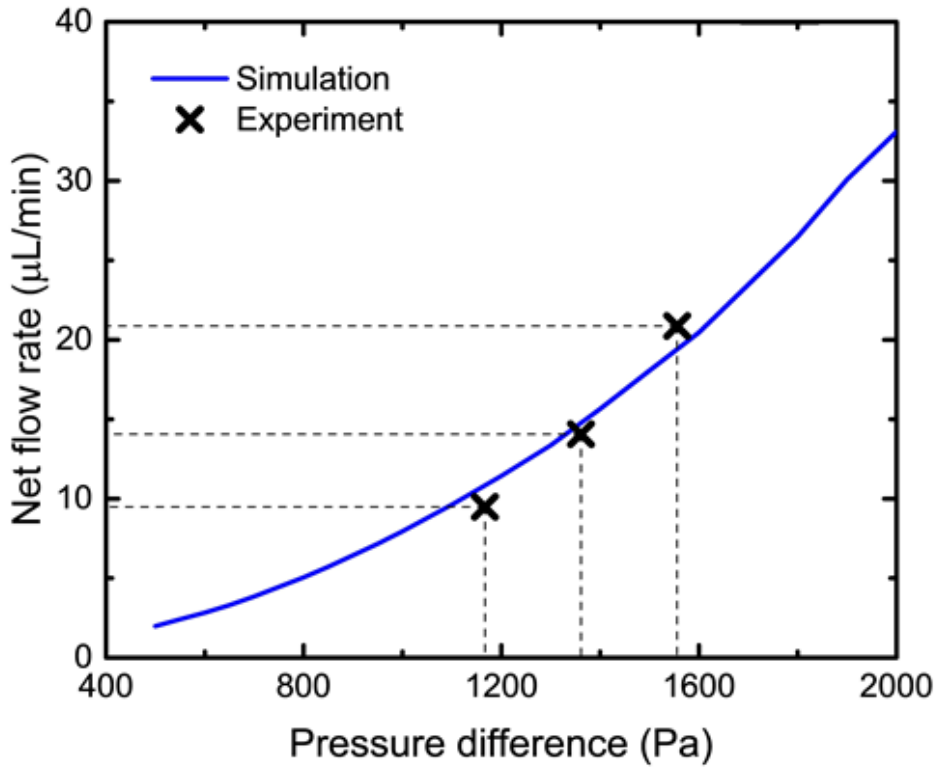

*Figure 2.122.Experimental flow rate compared to the FEA simulation results for the fabricated micropump.*

## 2.7    Concluding Remarks

In this work, a process for the fabrication of a piezoelectrically actuated diffuser micropump was presented, together with an FEA simulation of the flow rates for nozzle/diffuser micropumps. The micropump thus fabricated was characterized by measuring flow rate against various applied input voltages. It was established that a working micropump with a rated flow rate of 10 μL/min, at an applied voltage of 12 V, can be fabricated using the given process. The pump efficiency was calculated to be 1.642, while the rectification factor was calculated to be 12.33%. The chamber pressure was found to be in the range of 1.1 to 1.5 kPa using the flow rates to calculate the pressure in the chamber by fitting experimental results with the simulation.

The micropump fabrication process established in this work can be repeated in the future for various combinations of diffuser/nozzle dimensions. Further characterizations of the fabricated micropumps can be done by measuring backpressure and chamber deflection for the given voltage and frequency of the piezoelectric actuator.

Other materials can be used for fabrication purposes. Spun and Cured PMMA can be used as a structural material for pump design in place of PDMS. Silicon substrate can be used in place of the glass substrate to integrate the micropump with semiconductor electronics for lab-on-a-chip applications.

# Measurement of viscosity in real-time using pressure sensors

## 3.1  Background

Viscosity can be defined as the resistance or friction of a fluid when it is in motion. It is one of the most important and unique properties of chemical and biological fluids. By measuring viscosity, we can determine a large amount of information about the fluid. Measurement of viscosity has applications in industries like fossil fuel, biomedical, chemical, and so on. In the past few decades, there have been several developments to measure viscosity to test and monitor liquids like blood, mucus, lubricants, fuels, and others [57]–[60]. Further, by measuring the viscosity of a mixture of fluid, the ratio of individual components in the mixture can be determined, for example, the amount of medicine in blood, amount of water in oil [61]–[63]. To measure viscosity, many techniques have been reported such as falling ball, moving paddle, capillary force, resonating microtube, tuning fork, etc. [64]–[68]. In recent years, several MEMS-based viscometers have been developed using microfabrication and microfluidic technology[69]. Even though measuring viscosity is a very important task [70]–[72], its measurement in real-time is not trivial. However, there are many applications where continuous real-time measurement of viscosity can prove invaluable.

## 3.2 Related Work

Measurement of viscosity has been a very intriguing area of research for the past few decades. Due to its applications in the major areas like biomedical in disease detection, the precision of drug delivery, handling samples and its mixtures, etc., and in oil industries in differentiating oil and water, handling adulteration, etc. Analysis of blood viscosity has played a huge role in the detection of diseases like diabetics. By analyzing the serum viscosity of a diabetic patient, it was found that the viscosity of serum has been increased [73]. Recently, Duan et al. has optimized how diabetic detection can be multiplexed via droplet array [74]. Rand et al. have discussed the change in viscosity of human blood due to hypothermic and normothermic conditions [75]. In his paper, Doffin et al. measured the viscosity of the blood and plasma using the falling ball technique, i.e. by observing the fall time of a ball in a disposable syringe [76]. Smith et al. in their work have designed and modeled theoretically an oscillating MEMS-based viscometer that can measure for unadulterated human blood. The mentioned device uses an oscillating plate structure and dependence of damping ratio of squeeze film on to surrounding fluid to determine fluid viscosity [77]. Other than human blood, an IoT-based milk quality monitoring system to stop milk adulteration using a viscosity sensor was presented by Rajakumar et al. [78].

There have been many different kinds and methods that have been used to make viscometers over the past decades. A falling ball technique was used by Cho et al., where depending on the time intervals between successive ball dropping the viscosity of the concentrated solutions was measured [79]. Brand et al. reported a low-frequency viscosity sensor that can be used for real-time polarization monitoring. These micromachined membrane resonators detect the viscosity using its electrothermal excitation and transverse membrane vibrations which can be detected by piezoresistivity [80]. Kim and co-authors, in their work, showed for the first time, how to make a dual-capillary-tube viscometer by just using two liquid-height measurements instead of flowrate and pressure measurements like done before [81]. Another dynamic viscometer using resonating microtube was demonstrated by Sparks et al., the viscosity can be measured by checking the damping effect of the liquid, due to the motion of the resonating tube [82]. Riesch et al. demonstrated a micromachined viscosity sensor having a rectangular plate with four beams as resonating parts, and uses piezoresistive readout and Lorentz force excitation to detect viscosity [83]. A viscosity sensor was designed by Noel et al., by using the drag force of a liquid in a laminar flow [84]. Puchades et al. presented a viscosity sensor that is based on thermally actuated silicon diaphragm and damping of the surrounding liquid. The actuation happens due to resistive heaters and piezoresistive sensing [85]. Another well-known method of making a viscometer is by using tuning forks. Heinisch et al. presented a tuning fork viscometer where the tuning forks are electromagnetically driven [86]. Another type of technique was shown by Heintzmann et al. by using the oscillation drop technique for measuring the viscosity of melted metals [87].

Another very important area where the viscosity sensors get the most utilization is in the oil industry. There have been many ways viscometers have been used and made over the years. A very important property of engine oil to check its quality is its viscosity. Jakoby et al. in their contribution has used a micro acoustic viscosity sensor to check the viscosity of engine oil and its dependency on temperature [88]. Another acoustic wave-based viscosity sensor was presented for measuring oil viscosity in real-time by Durdag [89]. Toledo et al. demonstrated a viscosity sensor to measure the viscosity of oil/fuel with uses a quartz tuning fork and resonators to do so. In this structure there are two resonators, one is AIN based rectangular plate and the other is a tuning fork. The impedance can measure from the in-plane movement of two plates [90]. A droplet-based simple viscometer was presented by Yunzi et al., which is a continuous viscometer for water-in-oil and can measure viscosity in 10 seconds or less. The viscometer has a geometry such as it generates droplets under pressure, constantly [91].

Recently, Maha and co-authors have established a unique viscometer that depends on velocity to measure viscosity in real-time and without disturbing the flow of the liquid. Here PMMA based microchannel is attached to a flexible PDMS-based platform and this assembly is placed inside a pipe. This viscosity sensor does not need a pumping mechanism, adjustable to different pipe curvatures, and can send data wirelessly [92]. Getting inspired by this novel technique, we have done our analysis mathematically and with simulation by using COMSOL Multiphysics. After comparing both the result we have concluded that the system is very efficient. This work will be discussed in this chapter of the thesis.

## 3.3  System Structure

The structure of the proposed system is presented in Figure 3.1. The system consists of a narrow PMMA channel with capacitive pressure sensors fabricated using copper bilayers on a PDMS substrate. The PMMA channel has a height of 250 µm from the inside of the pipe, while the length of the channel is 60 mm. Such a narrow channel provides a laminar flow, thus allowing for stable measurement of pressure difference. The pressure sensors are placed at the ends of the channel to read the pressure continuously. When fluid passes through the pipe, it also enters the channel at the velocity-dependent on the total flow rate in the pipe. The fluid exerts pressure on the walls of the channel, which can be measured using the two pressure sensors. The difference in the pressure measurement is proportional to the flow rate (as in the case of Ohm's Law for the flow of electrons), with the constant of proportionality being the resistance of the channel. This resistance is a function of fluid viscosity, thus, for a given flow rate, the pressure difference in the channel can be used to measure the viscosity of the fluid.

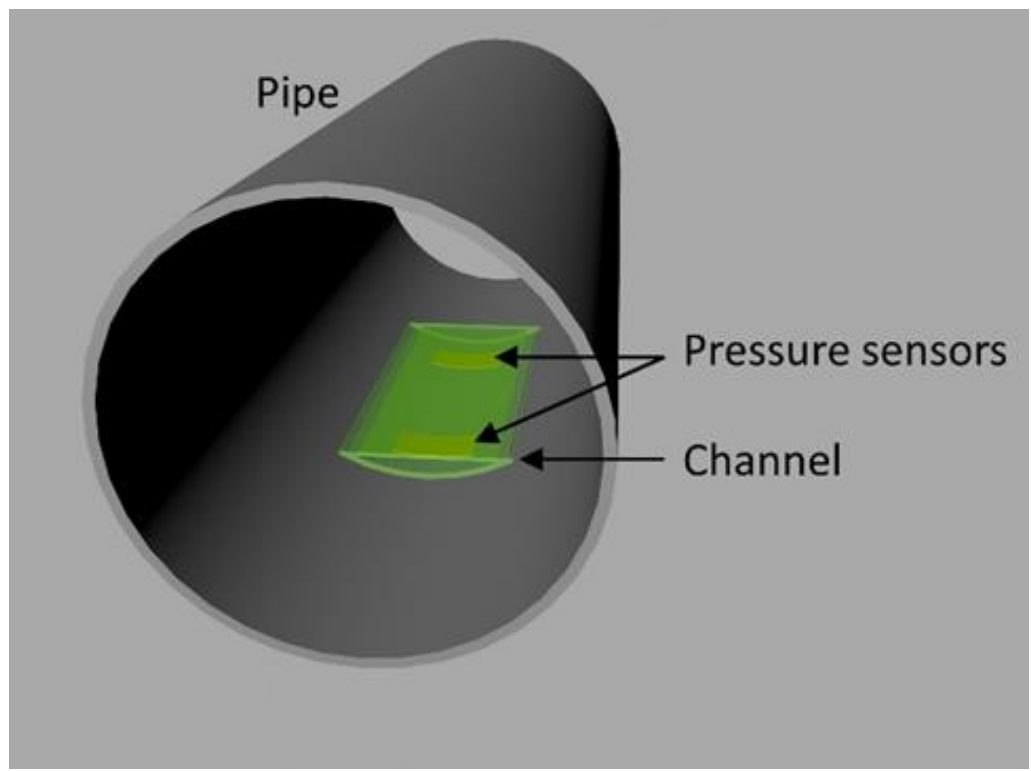

*Figure 3. 1 Simplified illustration of the proposed viscosity measurement system.*

## 3.4   Mathematical Modelling

To obtain the relationship between the pressure difference and fluid viscosity, we assume an incompressible, steady, laminar fluid of viscosity $\mu$ in cylindrical pipe of radius R. In the circular pipe, the fluid flowrate counters from concentric circles [71]. Thus, according to the Hagen-Poiseuille equation, derived from the Navier-Stokes equation, the volumetric flow rate of a thin circular fluid layer of thickness dr, at the distance r from the center, is given by:

$$Q(r)dr = -\frac{1}{4\mu}\frac{\Delta P}{L}(R^2 - r^2) \times 2\pi r dr$$

(3.1)

where $\Delta P$ is the pressure difference between two points at a distance L from each other. For the large circular pipe, the total flow rate in the pipe is given by:

$$Q_R = \int\limits_0^R Q(r)dr$$

(3.2)

Thus, from equation (3.1), we obtain the in the pipe flowrate as:

$$Q_R = -\frac{1}{4\mu}\frac{\Delta P}{L} \times 2\pi\left[\int_0^R (R^2 - r^2)\, dr\right] \tag{3.3}$$

$$Q_R = -\frac{1}{4\mu}\frac{\Delta P}{L}\left(\frac{\pi R^4}{2}\right) \tag{3.4}$$

For a channel placed at the inside boundary of the big circular pipe (Figure 3.1), the flow rate in the channel is given by:

$$Q_{R-h} = \left(\frac{\theta}{2\pi}\right)\int_{R-h}^{R} Q(r)\, dr \tag{3.5}$$

where h is the height of the channel above the surface of the pipe and θ is the sectoral angle of the channel. From equation (3.1), we have,

$$Q_{R-h} = -\frac{1}{4\mu}\frac{\Delta P}{L}\left[\frac{\pi R^2}{2}\right]\{R^2 - (R-h)^2\} \times \left[1 - \frac{1}{R^2}(R-h)^2\right] \tag{3.6}$$

In terms of the total flowrate inside the pipe, the flow in the channel is given by:

$$Q_{R-h} = \frac{Q_R}{R^2}\{R^2 - (R-h)^2)\} \times \left[1 - \frac{1}{R^2}(R-h)^2\right] \tag{3.7}$$

The average velocity in the channel of height can be obtained from the total flowrate using the relationship $\bar{v} = Q/A$ as:

$$\bar{v}_{R-h} = \frac{Q_R}{\pi R^2}\left[1 - \frac{1}{R^2}(R-h)^2\right] \tag{3.8}$$

From reference [72], we know that the average velocity in a tubular channel of height h is given by:

$$\bar{v}_{R-h} = \frac{\Delta p R^2}{8\mu L}\left[1 + \left(\frac{R-h}{R}\right)^2 + \frac{\left(\frac{R-h}{R}\right)^2 - 1}{\ln\left(\frac{R}{R-h}\right)}\right] \tag{3.9}$$

Equations (3.8) and (3.9) represent two different ways of expressing the same quantity. Hence, we can equate these two equations, and solve for the viscosity of the fluid in terms of the other parameters:

$$\mu = \frac{\pi R^4 \Delta p \left[1 + \left(\frac{R-h}{R}\right)^2 + \frac{\left(\frac{R-h}{R}\right)^2 - 1}{\ln\left(\frac{R}{R-h}\right)}\right]}{8LQ_R\left[1 - \left(\frac{R-h}{R}\right)^2\right]} \tag{3.10}$$

Thus, if pressure and flow rate through a pipe is measured continuously, we can obtain the viscosity of the fluid in real-time.

## 3.5   Simulation & Analysis

We performed finite element analysis to validate our mathematical model. The simulation was done using COMSOL Multiphysics. The geometry used for simulation consisted of a circular pipe of radius 19 mm and length 60 mm. A channel with 0.25 mm height and 3 mm width was inserted at the outer radius of the pipe (Figure 3.2). We measured the pressure at the two ends of the channel for the simulation, we fixed the flowrate in the pipe to be 1500 ml/min. The velocity distribution in the pipe and the channel are as shown in Figure 3, for a liquid mixture of 85% glycol and 15% water. In the pipe, the maximum velocity is observed at the center, while the velocity is zero at the pipe surface. The distribution of velocity varies with the distance from the center (as expected), while inside the channel, the velocity distribution is more complex. The average velocity in the channel is found to be consistent with the expected value from the mathematical model (Equation 3.8).

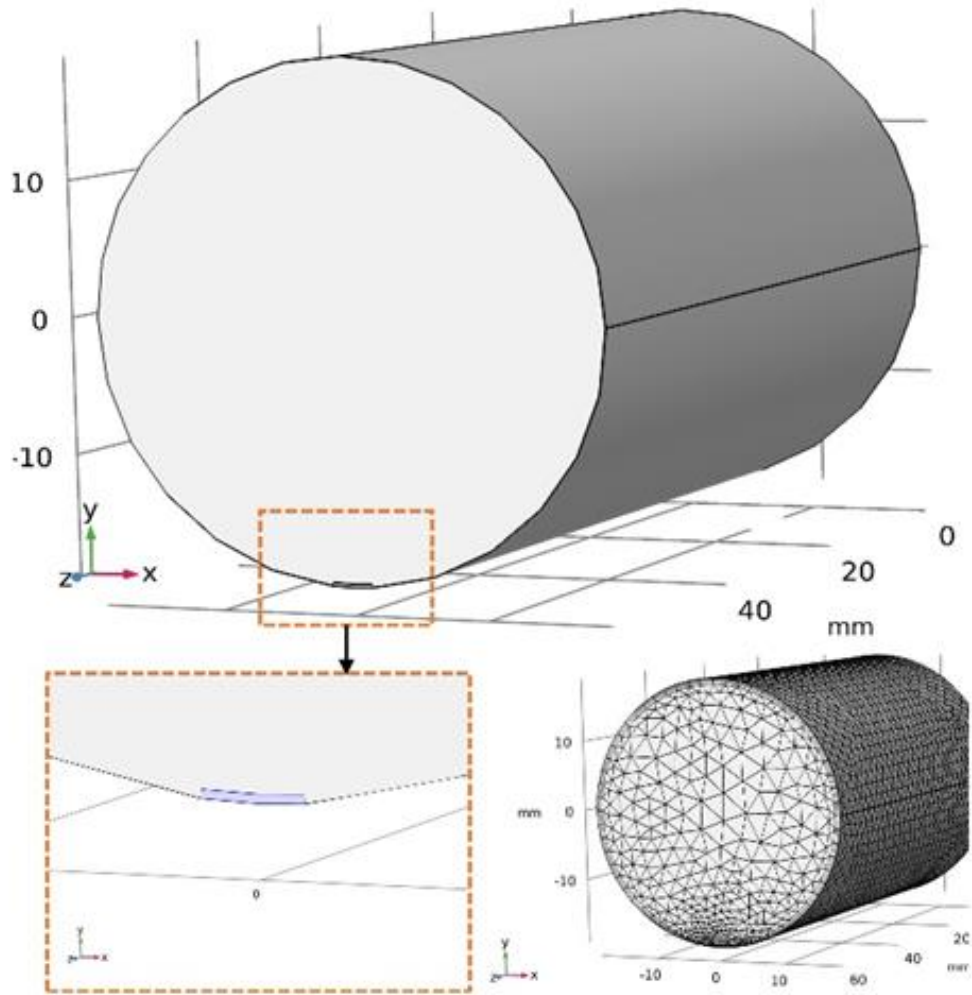

*Figure 3. 2 Geometry of the simulated system with the channel inserted in the pipe. The meshing
is done such that the detailing of the tiny channel is captured in the simulation.*

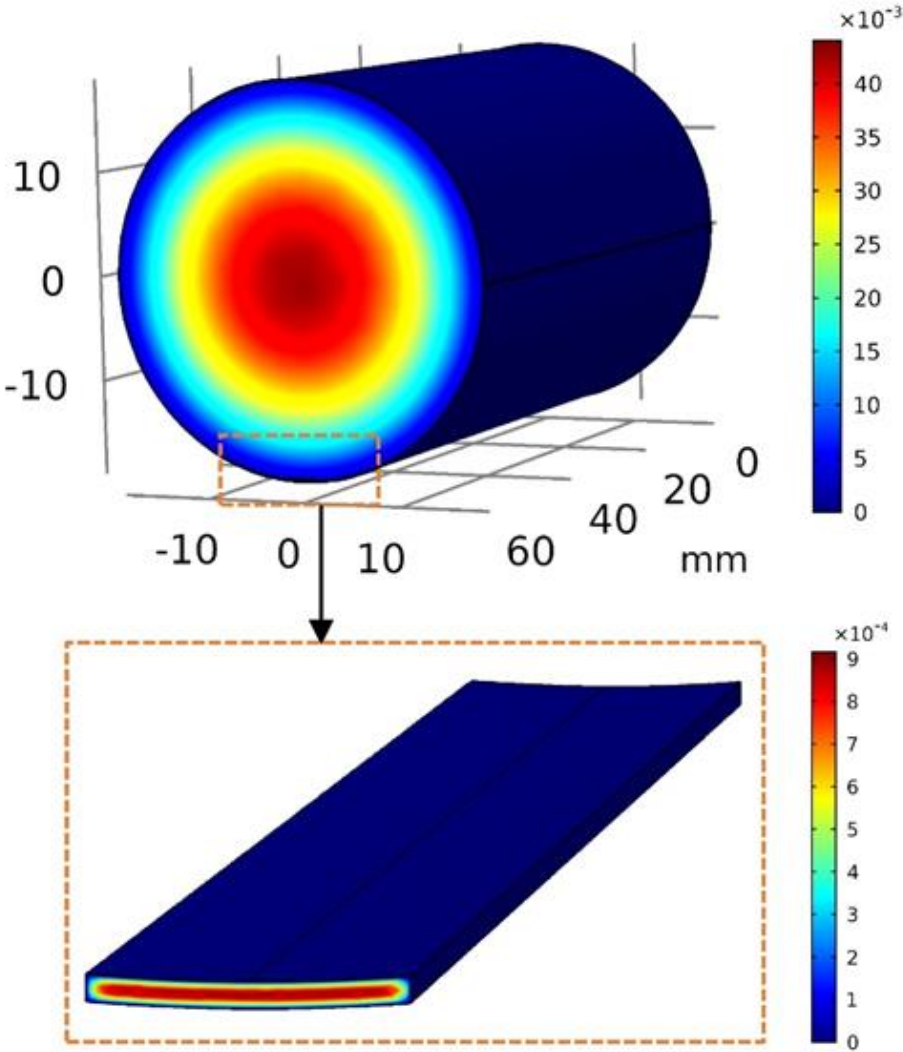

*Figure 3. 3 Distribution of fluidic velocity in the cross section of the pipe and the channel for flowrate of 1500 ml/min and liquid viscosity of 0.13 Pa-s. The color scales are in m/s.*

The key result is the distribution of pressure inside the channel as a function of the distance. As seen from Figure 3.4, the pressure is higher close to the inlet of the fluid and reduces to zero (relatively) at the outlet. In this case, the maximum pressure in the channel was observed to be 895 Pa. While the pressure varies from higher to lower when going from inlet to outlet, the velocity at each point in the channel remains constant in the steady state.

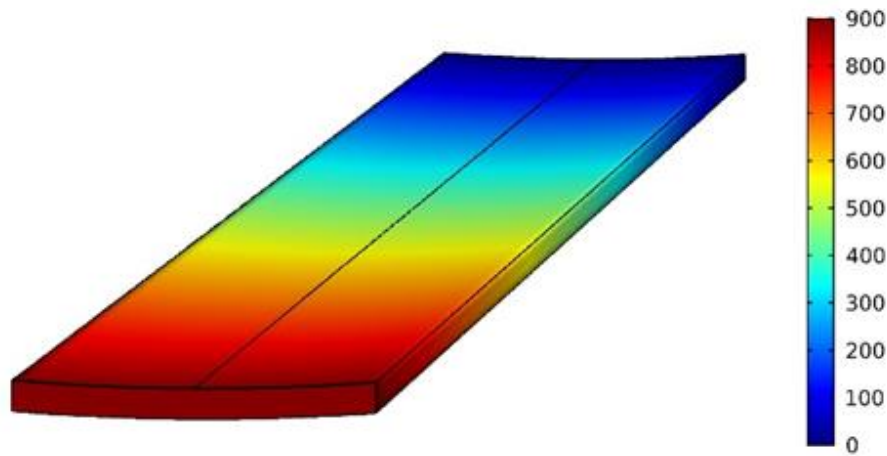

*Figure 3. 4 Relative pressure distribution in the channel of height 0.25 mm placed along the outer diameter of the pipe. The color scale is in Pascals.*

## 3.6  Results Obtained

In our mathematical analysis, we found that the pressure difference along the channel depends on the geometry of the channel and pipe (length and height of the channel, and the radius of the pipe), the flow rate, and the viscosity of the fluid. Given that the geometry of a system is fixed for a particular deployment, we compared the results of the simulation by varying the viscosity of the fluid to observe the pressure drop across the channel. If the predicted pressure drop is achieved, the same equation can be used to estimate viscosity given a known flow rate and pressure difference. The viscosity of the simulated fluid was varied between that of water and glycol by increasing the percentage of glycol in water in steps of 5% (to eventually reach 100% glycol). Figure 3.5 shows the variation of the observed pressure difference with fluids of various viscosities. The symbols represent results from the simulation, while the line represents the analytical results obtained from Equation (3.10). These results compare favorably.

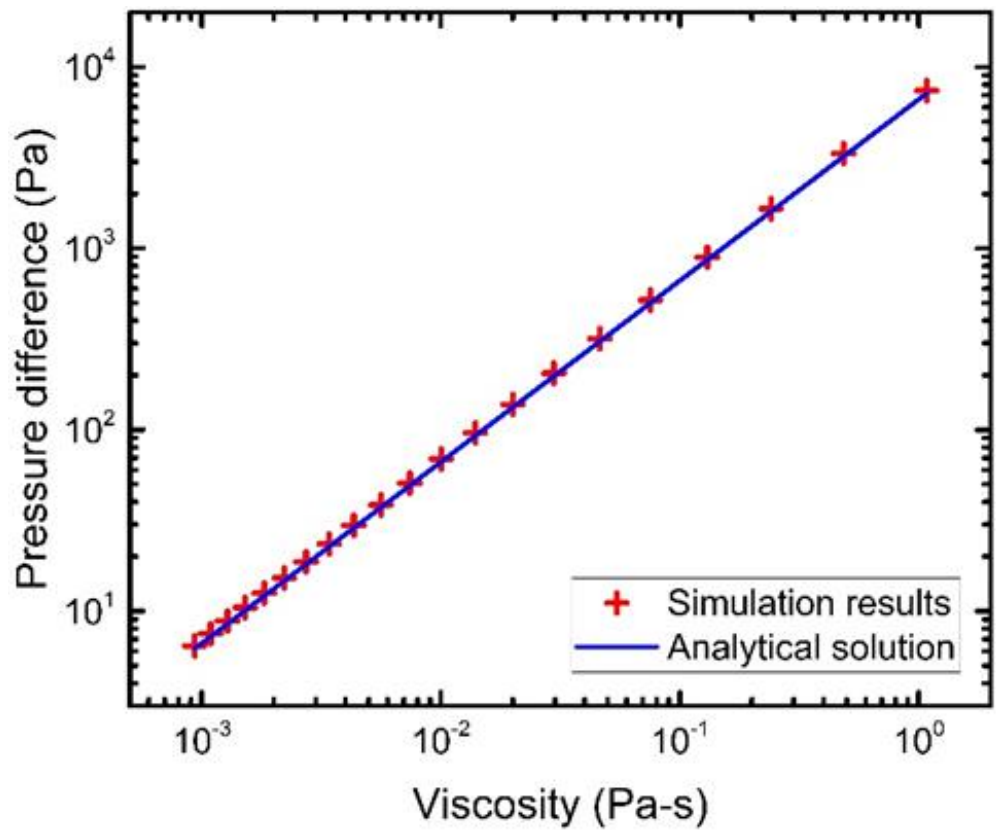

*Figure 3. 5 Change in pressure-difference in the channel with change in viscosity of fluid. The simulation results (symbols) match closely with those obtained from the mathematical model (line).*

As seen from the simulation (and mathematical model), the pressure difference inside the channel changes linearly with fluid viscosity. Thus, the pressure difference is lowest for 100% water, while is it highest for 100% glycol.

## 3.7    Concluding Remarks

In this paper, we presented a novel approach to the measurement of fluid viscosity in real-time. A flow meter and a pair of pressure sensors are used to create the setup for viscosity measurement. The proposed system is encapsulated in a PMMA channel to obtain laminar flow and stable pressure drop across the system. We carried out mathematical modeling of the system to determine the method of obtaining viscosity given fluid flow rate and the pressure difference in the channel. Our mathematical model was verified using finite element analysis (FEA) simulation of the system. The results from both analyses match closely. Thus, we can conclude that viscosity can be calculated if we can monitor the change in pressure due to the liquid inside the pipe along with the flow rate continuously. This measurement of viscosity in real-time can be applied to help determine the amount of oil in water in the petroleum industry, the ratio of liquids in a solution in the chemical industry, the amount of medicine in blood in the medical industry, and so on.

# Current Work: 3D Printed Low-cost Fully Flexible Transparent Nozzle/Diffuser Mini-pump

With the advent of miniaturized electronic devices and wearable electronics, miniature laboratories made it possible to analyze bio-fluids instantly [92]–[98]. Being one of the most important parts of a microfluidic system, the pump helps in inserting medicine into the body, exerting bodily fluids, controlling fluid flow, mixing two or more chemicals, sample separation, etc. The pump designs we get in the literature are mostly fabricated in a cleanroom facility, which makes the process expensive and strenuous. As a result, the device becomes unaffordable for the masses and can be too fragile to handle. On the other hand, if we can use a simple 3D printer, the steps of fabrication reduces, can be made in bulk, and the cost gets cut down tremendously. Nowadays 3D printed systems are being very popular in biomedical applications due to it's being convenient in handling, and cost-effectiveness. A 3D printer can handle complex designs easily and it can be manufactured from any kind of setup, small or big. In this work, we are reporting a fully functional 3D printed novel mini-pump which was made without using a cleanroom facility.

In our previous work, we have shown a miniaturized nozzle/diffuser micro-pump, to fabricate that we needed to use cleanroom facilities. In this work, we are presenting a 3D printed, fully flexible, functionally accurate mini-pump. We assembled the entire system using PDMS as it is a biocompatible, flexible, and transparent material, which is a great choice for biomedical applications. To set the pump up we have 3D printed a mold with a thickness of 7mm and the trapezoidal diode has a width of 4mm on one side and 20mm on the other. The circular chamber has a diameter of 300mm. The mold is created using white PLA as printing material. The entire system was fabricated using PDMS, curated at 10:1 ratio, and was poured over the mold. A black acrylic rectangular base was used to hold the shape while the PDMS gets dried up overnight instead of using a hotplate. Fig 4.1 shows PDMS kept on the mold overnight. Once the PDMS got baked the structure was removed from the acrylic holder and the mold was detached from the sensor. A separate acrylic rectangular structure was used to give shape to the PDMS bottom to hold the pump structure. The mold imprinted PDMS structure was attached to the bottom PDMS using external pressure and the adhesiveness of PDMS and it was made sure there is no air bubble left where the top PDMS and bottom PDMS made connections. In Fig 4.2 we can see the fully fabricated minipump.

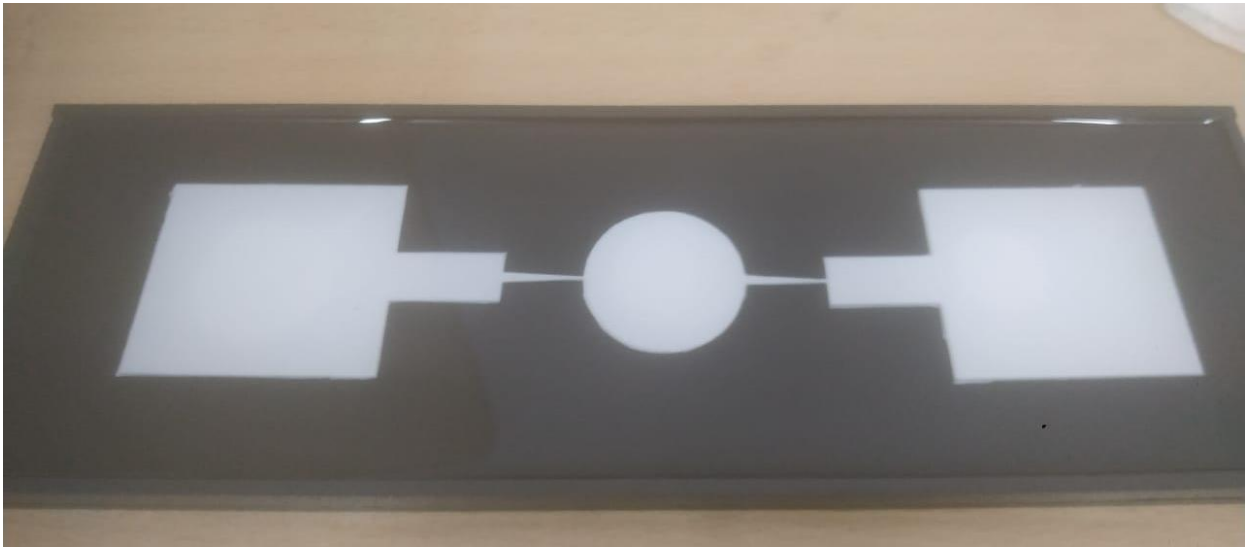

*Figure 4. 1. An intermediate stage for making a fully flexible microfluidic pump*

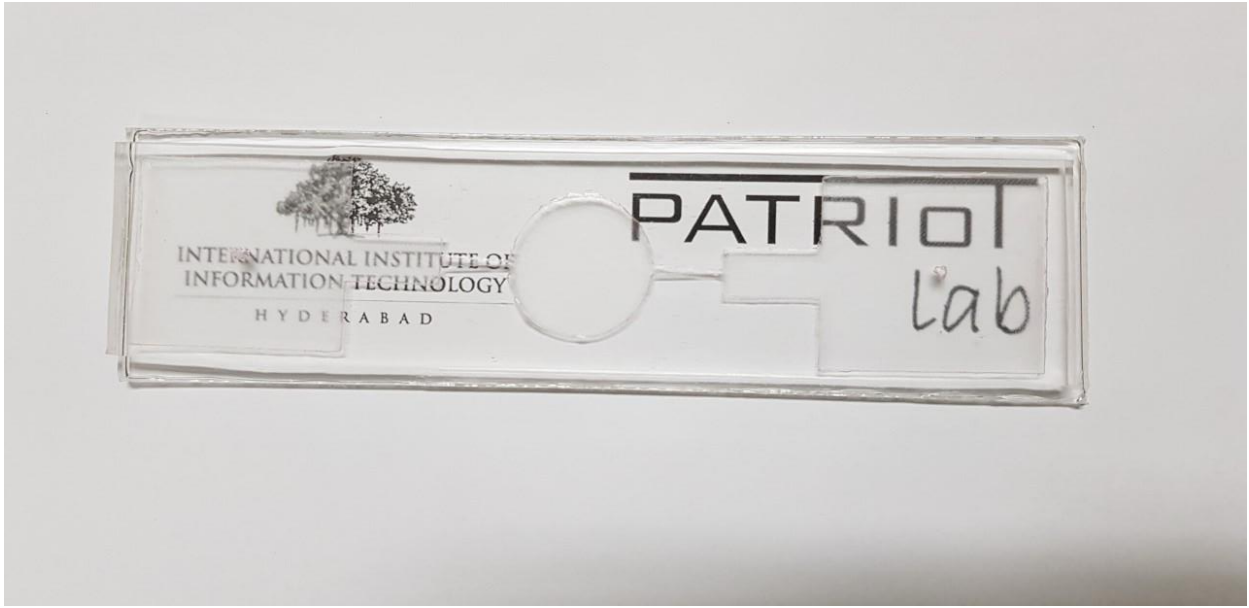

*Figure 4. 2. Fully flexible transparent minipump*

To test the mini-pump we again 3D printed a pressure device controlled by an Arduino and dc motor. The mini-pump was tested by applying pressure on the circular chamber and we can gladly report that the pump is fully functional. The amount of liquid that can be travel inside the pump was checked, and also two or more chemicals mixing inside the pump were observed. The result is very promising. The see-through nature of the entire system helped to notice the working of the mini-pump and the flexible nature of the pump is best suited for wearable electronics[100]–[104]. This novel mini-pump can bring a lot of opportunities in biomedical researches like examining samples, mixing chemicals, inserting medicines, etc.

# Conclusion & Future Work

In this thesis, we have discussed two microfluidic sensors, one for biomedical applications and the other for mainly industrial usage. After analyzing by simulating and fabricating, we can conclude, both the devices are functionally efficient and structurally simple.

In the first section, we have designed a microfluidic micropump, which has been simulated, fabricated, and experimentally analyzed. The change in flow rate with applied voltage was presented. Successful fabrication of a micropump which have a flow rate of 10 μL/min at an applied voltage of 12V was done [105]. We can conclude by this experiment that this simple but efficient structure can be replicated with various materials.

In the second part of the thesis, we have discussed a viscosity sensor that utilizes the flow rate of the liquid using pressure sensors. The simulation results were supported by an elaborate mathematical analysis. From that, we can conclude if the device is to be made, it will work very efficiently. The viscosity sensor can be made using various other material which is to be explored.

The design of the micropump has a significant contribution to biomedical researches. To mix two types of chemicals for testing samples in lab-on-chip, this design will be very useful. The viscosity sensor solves a big problem in the oil industry where the monitoring of the oil quality is essential.

# Publications

## • Related Publications

1. **S. Bhattacharjee**, R. B. Mishra, D. Devendra, A. M. Hussain "Simulation and Fabrication of Piezoelectrically Actuated Nozzle/Diffuser Micropump" IEEE Sensors 2019 (*Accepted*)

2. **S. Bhattacharjee**, R. B. Mishra, S. Malkurthi, A. M. Hussain "Measurement of viscosity in real-time using pressure sensors" IEEE Sensors 2021 (Submitted)

3. **S. Bhattacharjee**, A. M. Hussain "3D Printed Low-cost Fully Flexible Transparent Nozzle/Diffuser Minipump" (Abstract Submitted)

## • Other Publications

1. W. Babatain, **S. Bhattacharjee**, A. M. Hussain, M. M. Hussain "Acceleration Sensors: Sensing Mechanisms, Emerging Fabrication Strategies, Materials, and Applications" ACS Applied Electronic Material (*Accepted*) [Cover Article]

2. K. K. S. Charan, S. R. Nagireddy, **S. Bhattacharjee**, A. M. Hussain "Design of Heating Coils Based on Space-Filling Fractal Curves for Highly Uniform Temperature Distribution", MRS Advances 2020 (*Accepted*)

3. R. B. Mishra, S. R. Nagireddy, **S. Bhattacharjee**, A. M. Hussain "Modelling and Simulation of Elliptical Capacitive Pressure Sensor", 2019 IEEE International Conference on Modelling of Systems Circuits and Devices (MOS-AK India, 2019) (*Accepted*)

# Patents

1. United States Patent and Trademark Office (USPTO) patent application No. 63/013,197 "Sustainable Design of Ventilator" Muhammad Mustafa Hussain, Sherjeel Khan, Sohail Shaikh, Nadeem Qaiser, **Sumana Bhattacharjee**